\documentclass[%
 reprint,
 amsmath,amssymb,
 aps,
 pra,
 floatfix
]{revtex4-2}

\usepackage{textcomp,gensymb}
\usepackage{braket}
\usepackage[normalem]{ulem}
\usepackage{graphicx}
\usepackage{dcolumn}
\usepackage{bm}
\usepackage{upgreek}
\usepackage{soul}
\usepackage{units}
\usepackage{multirow}

\usepackage[export]{adjustbox}

\newcommand\thefont{\expandafter\string\the\font}
\usepackage{hyperref}

\begin{document}

\title{Long distance optical conveyor-belt transport of ultracold $^{133}$Cs and $^{87}$Rb atoms}

\author{Alex J. Matthies}
\author{Jonathan M. Mortlock}
\author{Lewis A. McArd}
\author{Adarsh P. Raghuram}
\author{Andrew D. Innes}
\author{Philip D. Gregory}
\author{Sarah L. Bromley}
\author{Simon L. Cornish}
 \email{s.l.cornish@durham.ac.uk}
\address{
\mbox{Department of Physics, Durham University, South Road, Durham DH1 3LE, United Kingdom}}

\begin{abstract}
We report on the transport of ultracold cesium and rubidium atoms over $37.2$\,cm in under $25$\,ms using an optical conveyor belt formed by two counter-propagating beams with a controllable frequency difference that generate a movable optical lattice. By carefully selecting the waists and focus positions, we are able to use two static Gaussian beams for the transport, avoiding the need for a Bessel beam or vari-focus lenses. We characterize the transport efficiency for both species, including a comparison of different transport trajectories, gaining insight into the loss mechanisms and finding the minimum jerk trajectory to be optimum. Using the optimized parameters, we are able to transport up to $7 \times 10^6$ cesium or rubidium atoms with an efficiency up to $75$\,\%. To demonstrate the viability of our transport scheme for experiments employing  quantum gas microscopy, we produce Bose-Einstein condensates of either species after transport and present measurements of the simultaneous transport of both species.
\end{abstract}

\maketitle

\section{Introduction}
Experiments utilizing ultracold mixtures of different atomic species are of great interest due to the rich interplay between intra- and inter-species interactions~\cite{Hadzibabic2002, Silber2005, Tung2014, Maier2015, Wang2015, Cetina2016, Ulmanis2016, Groebner2016, Johansen2017, DeSalvo2017, Reichsoellner2017, Burchianti2018, DeSalvo2019}, and for the formation of ultracold molecular gases~\cite{Ni2008, Takekoshi2014, Molony2014, Park2015, Molony2016, Guo2016, Rvachov2017,  Seesselberg2018, Voges2020, Rosenberg2022, Stevenson2023}. In single species experiments, the use of multi-chamber setups where the atoms are transported from one chamber to another using optical or magnetic traps has allowed for experiments with ever-increasing complexity. For example, this has enabled the investigation of quantum gas microscopy~\cite{Bakr2009, Sherson2010, Cheuk2015, Edge2015, Haller2015, Parsons2015, Omran2015, Yamamoto2016}, cavity QED~\cite{Sauer2004, Hickman2020}, hollow-core fibers~\cite{Okaba2014, Langbecker2018} and nanofibers~\cite{Schneeweiss2013}. The development of transport techniques that are compatible with the different properties and requirements of multiple atomic species are important for the continued development of ultracold mixture and molecule experiments. Increased experimental complexity often leads to longer duty cycles and hence drives the the need for fast and efficient transport.

A variety of atomic transport schemes have been developed and implemented, each with their advantages and limitations. For atoms with magnetically trappable states that are stable against inelastic collisions, a mechanically moving coil pair~\cite{Lewandowski2002, Nakagawa2005, Pertot2009, Haendel2011} or a series of overlapping coils with time-varying currents~\cite{Greiner2001, Haensel2001, Minniberger2014} can be used. These schemes have been demonstrated to work over long distances with minimal heating~\cite{Greiner2001} and over short distances for precise positional control~\cite{Haensel2001}. Magnetic traps have the advantage of a larger trap depth and volume, however they can take up a significant amount of optical access and often require transport times of several seconds.

Optical transport schemes can be used for all atomic species, independent of internal ground state. They require significantly less space than magnetic transport schemes, making them usable for a much greater range of experiments. The most straightforward method uses a translation stage to move the focus position of a single Gaussian beam optical dipole trap~\cite{Gustavson2001, Couvert2008, Naides2013}. However the translation stage can be a significant source of mechanical vibrations and this scheme suffers from weak axial confinement along the propagation direction, and hence transport direction. This leads to elongated atomic clouds and long transport times. Variations of the scheme have also been demonstrated, including passing two beams through the translated lens to form a shallow-angle crossed optical dipole trap~\cite{Gross2016} or a pair of translation stages to shift the position of a lattice~\cite{Middelmann2012}. Using a variable-focus lens: either an electrically tunable liquid-based lens~\cite{Leonard2014} or a phase-pattern Moir\'e lens~\cite{Unnikrishnan2021}, eliminates noise due to mechanically moving components but still suffers from weak axial confinement. Novel hybrid optical and magnetic transport schemes have also been developed \cite{Pritchard2006, Marchant2011}.

A different approach is to use an optical conveyor-belt scheme, where a moving lattice is used to transport the atoms~\cite{Schrader2001, Schmid2006, Klostermann2022, Bao2022, TrisnadiQuantumMatterSynthesizer2022}. Here, the use of an optical lattice eliminates the problem of weak axial confinement. By introducing a frequency difference $\Delta f$ the lattice is translated at speed $v =  \lambda \Delta f /2$ along the propagation direction of the beam with the greater frequency. Dynamically changing the detuning between the lattice beams allows the translation of the lattice sites to transport the ultracold sample. Typically transport over $\sim30$~cm can be achieved in less than 50~ms, and the initial axial extent of the cloud is preserved by the lattice. 

Our ambition is to develop a quantum gas microscope for RbCs molecules incorporating fast and efficient optical transport, as illustrated in Fig.\,\ref{fig:radialComponents}(a). Imaging atoms in an optical lattice with single-site resolution requires a large-NA objective lens sited close to the atoms, in our case below a glass ``science cell''. To maintain the required optical access in the science cell atoms are initially prepared using a magneto-optical trap (MOT) in a different region of the vacuum apparatus (the ``MOT chamber'') and must therefore be transported $37.2$\,cm to the science cell. As fast duty cycles are desirable, we employ degenerate Raman sideband cooling (DRSC) in the MOT chamber following the initial laser cooling and hence we wish to quickly transport samples of $^{133}$Cs and $^{87}$Rb, hereafter referred to as Cs and Rb respectively, with temperatures of a few $\mu$K. Optical conveyor-belt transport is thus ideal for our needs, with short transport times and comparatively easy spatial mode-matching to the 3D optical lattice used in the quantum gas microscope. Here we show that efficient transport of Cs and Rb over $37.2$\,cm, both separately and simultaneously, is possible using an optical conveyor belt formed by two Gaussian beams with displaced focuses, aided by magnetic levitation at the start and end of the transport path. 

The structure of the paper is as follows. The theoretical design of our transport scheme is presented in Sec.\,\ref{sec:theory} where we establish the optimum beam waists and focus positions by calculating the trap depth along the transport path. In Sec.\,\ref{sec:expSetup} we review our experimental setup and procedure used to cool and trap the atoms in the MOT chamber prior to transport. The transport for both Cs and Rb is characterized and optimized separately in Sec.\,\ref{sec:transportCharacterize}, including a systematic investigation into different transport trajectories for Cs. To demonstrate the effectiveness of our transport scheme, in Sec.\,\ref{sec:BEC} we present results for the production of Bose-Einstein condensates of either species in the science cell. Finally, in Sec\,\ref{sec:DualSpecies} we demonstrate simultaneous transport of Cs and Rb without any decrease in efficiency.

\section{Transport Calculations\label{sec:theory}}

\begin{figure}[!ht]
\includegraphics[width=\linewidth]{"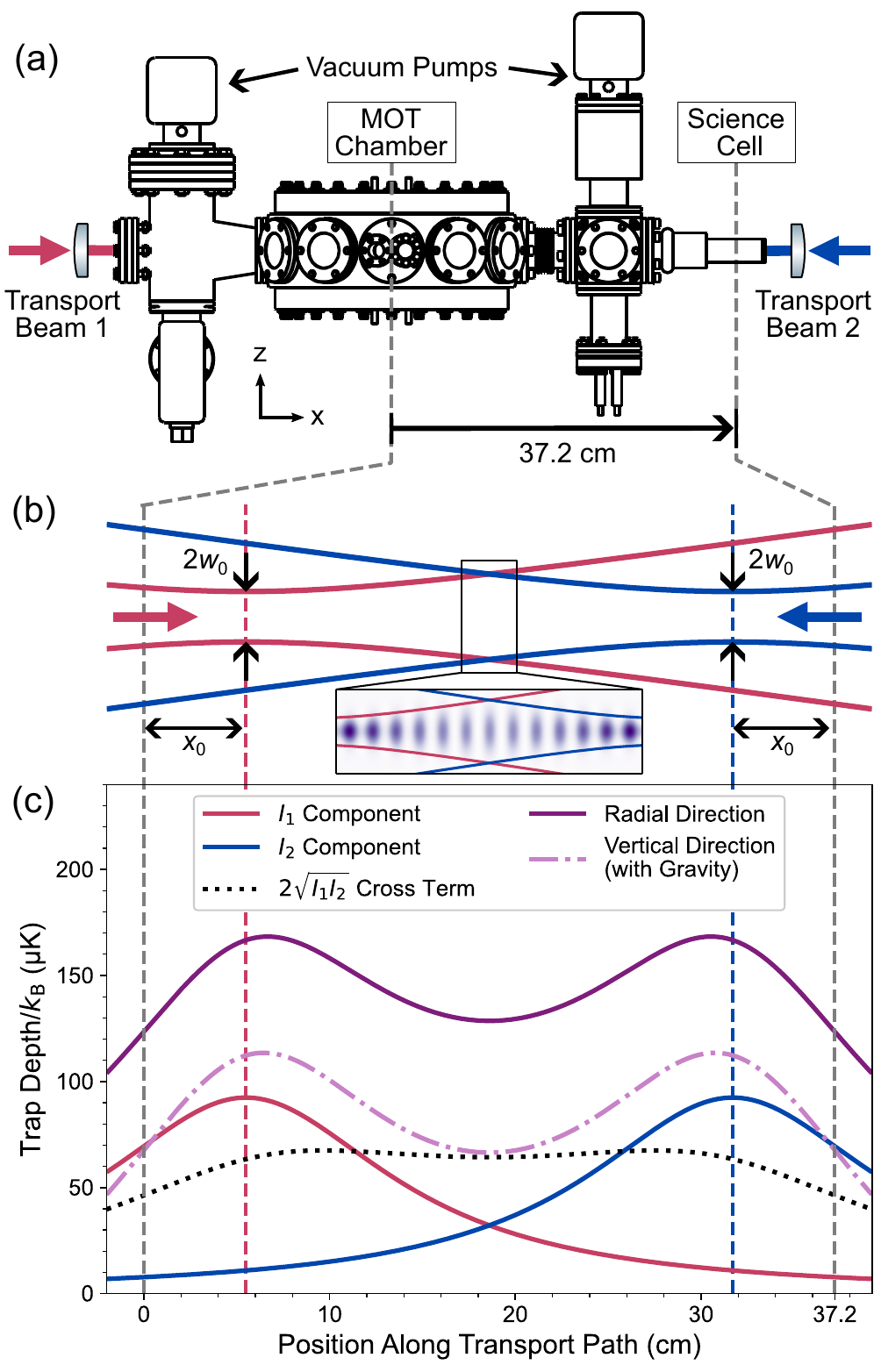"}
\caption{\label{fig:radialComponents} 
Overview of the transport problem and solution.
(a) Schematic of the vacuum apparatus. Transport is performed between the ``MOT chamber'' where the atoms are initially cooled and the ``science cell'' where future experiments will be performed. 
(b) Sketch of beam layout. Due to the symmetry of the problem, waists $w_0$ and focus positions $x_0$ of each beam are taken to be equal, with transport beam 1 (beam 2) focused closer to the MOT chamber (science cell). The inset shows a cartoon of the lattice produced by the interference of the beams with the varying beam sizes along transport path. Note that for clarity the radial divergence has been greatly exaggerated relative to the lattice spacing. 
(c) Contributions to the total radial trap depth (purple solid line) from beam 1 (red solid line), beam 2 (blue solid line) and the interference term (black dotted line). In the vertical direction the trap depth is reduced due to gravity as shown by the magenta dash-dotted line. The trap depths are calculated for Cs with $18$\,W of 1064\,nm light in each beam, $x_{\text{0}} = 5.5$\,cm and $w_{\text{0}} = 180\,\mu$m. 
}
\end{figure}

The challenge for optical conveyor belts is maintaining sufficient trap depth throughout the whole of the transport path. This stems from the radial divergence of Gaussian beams, leading to reduced trap depth away from the foci of the beams. This problem is exacerbated by gravity which ``tilts'' the trapping potential and thereby reduces the effective trap depth in the vertical direction. One solution is to use Bessel beams, as they have a central order with no radial divergence~\cite{Schmid2006, Klostermann2022}. However, a good quality Bessel beam can be hard to create, requiring a high quality axicon lens. Additionally, the fraction of the power into the axicon that is carried by the desired order is often only $\sim 10$\,\%, especially as the distance from the axicon increases~\cite{Klostermann2022}. Another solution is using both a moving lattice and a variable-focus lens~\cite{Bao2022}. However, we show below that these techniques can be avoided by carefully selecting the waists $w_0$ and focus positions $x_0$ of two Gaussian beams with equal powers used to form an optical conveyor belt, as shown in Fig.\,\ref{fig:radialComponents}. This approach, combined with magnetic levitation at both ends of the transport path, provides sufficient trap depth for efficient transport. 

The potential of an optical dipole trap is given by~\cite{Grimm2000}
\begin{equation}
\label{eqn:opticalDepth}
U(\mathbf{r}) = -\frac{\alpha_{\mathrm{\lambda}}}{2\epsilon_0 c}I(\mathbf{r}),
\end{equation}
where $\alpha_{\mathrm{\lambda}}$ is the real part of the atomic polarizability and $I(\mathbf{r})$ is the position dependent intensity. The polarizability is dependent on the atomic species, as well as the wavelength of light used. Our transport uses light at $1064$\,nm. At this wavelength the polarizabilities in atomic units~\footnote{In atomic units $4 \pi \epsilon_0 = 1$. Therefore in SI units the polarizabilities are $1162\times 4 \pi \epsilon_0 {a_0}^3$ for Cs and $687 \times 4 \pi \epsilon_0{a_0}^3$ for Rb} are $1162\,{a_0}^3$ for Cs and $687\,{a_0}^3$ for Rb~\cite{Safronova2006}. In the following sections, where a trap depth is shown, the values for Cs are used but all calculations were performed for both species. While the two species are different in terms of their polarizabilities, masses and ground state magnetic moments, these differences scale the magnitude of the trap depth but do not affect the optimum beam parameters.

The intensity of a Gaussian beam propagating along the $x$ axis, as a function of position, is given by
\begin{equation}
\label{eqn:gaussianI}
I(x,y,z) = \mathcal{I}(x) \exp{\left[- \frac{2 \left(y^2 + z^2\right)}{{w(x)}^2}\right]},
\end{equation}
where $\mathcal{I}(x)$ is the peak intensity at position $x$ and $w(x)$ is the position-dependent beam size. Respectively, they are given by
\begin{equation}
\label{eqn:peakI}
\mathcal{I}(x) = \frac{2 P}{\pi} \frac{1}{w(x)^2},
\end{equation}
and
\begin{equation}
\label{eqn:beamShape}
w(x) = w_{\mathrm{0}}\sqrt{1+\left( \frac{x - x_{\mathrm{f}}}{x_{\mathrm{R}}} \right)^2}, 
\end{equation}
where $P$ is the beam power, $w_0$ is the waist, $x_f$ is the focus position and $x_{\mathrm{R}}$ is the Rayleigh range. 

For conveyor-belt transport we have two beams counter-propagating along $x$ with intensities $I_1$ and $I_2$. They produce a standing wave interference pattern with an intensity varying according to~\cite{Hecht2002_Book}
\begin{equation}
\label{eqn:latticeI}
I(x,y,z) = I_{\mathrm{1}} + I_{\mathrm{2}} + 2\sqrt{I_{\mathrm{1}} I_{\mathrm{2}}}\cos\left(2kx \right).
\end{equation}
Here the coordinate dependence of $I_1$ and $I_2$ has been omitted. Combining the above equations gives the total optical potential, from which the trap depths can be found. Axially, along the direction of beam propagation, the part of the potential relevant for our transport is sinusoidal~\footnote{The overall potential is sinusoidal on top of a background of intensity $I_{\text{min}} = I_{\text{1}} + I_{\text{2}} - 2 \sqrt{I_{\text{1}}I_{\text{2}}}$. However any atoms trapped only by this background and not by the sinusoidal lattice, can not be accelerated/decelerated. Hence they are considered lost and the trap depth given only by the sinusoidal part of the potential.} with the amplitude determining the trap depth
\begin{equation}
\label{eqn:depthAxial}
\mathcal{U}_{\mathrm{axial}} = \frac{2\alpha_{\mathrm{\lambda}}}{\epsilon_{\mathrm{0}} c}\sqrt{\mathcal{I}_{\mathrm{1}}\mathcal{I}_{\mathrm{2}}}.
\end{equation}
Radially, perpendicular to the direction of beam propagation, the optical potential has the typical Gaussian shape and hence the trap depth is given by the peak intensity at the standing wave anti-nodes
\begin{equation}
\label{eqn:depthRadial}
\mathcal{U}_{\mathrm{radial}} = \frac{\alpha_{\mathrm{\lambda}}}{2\epsilon_0 c} \left( \mathcal{I}_{\mathrm{1}} + \mathcal{I}_{\mathrm{2}} + 2\sqrt{\mathcal{I}_{\mathrm{1}}\mathcal{I}_{\mathrm{2}}}\right).
\end{equation}
However, due to the effect of gravity the radial trap depths are different in the horizontal and vertical directions. Horizontally, the trap is purely Gaussian with a depth given by Eq.\,\ref{eqn:depthRadial}. Vertically there is an additional gravitational term $mgz$ which ``tilts'' the potential, lowering the trap depth. Here $m$ is the atomic mass and $g$ is the acceleration due to gravity. There is no straightforward analytic expression for the reduction in trap depth so the vertical trap depth is found by calculating the potential along the $z$ axis at each point along the transport path and numerically finding the difference between the local maximum and local minimum in the trap potential. Fig.\,\ref{fig:radialComponents}(c) shows the contributions to the radial trap depth from the three terms in Eq.\,\ref{eqn:depthRadial}, as well as the drop in trap depth due to the gravitational tilt, for two $1064$\,nm beams each with $18$\,W of power, waists of $w_{\text{0}} = 180\,\mu$m and focused a distance $x_{\text{0}} = 5.5$\,cm from their respective ends of the transport path. We show below in Sec\,\ref{sec:beamShapes} that these parameters optimize the trap depth along the transport axis.

\subsection{Optimizing the beam parameters \label{sec:beamShapes}}

\begin{figure} 
\includegraphics[width=\linewidth]{"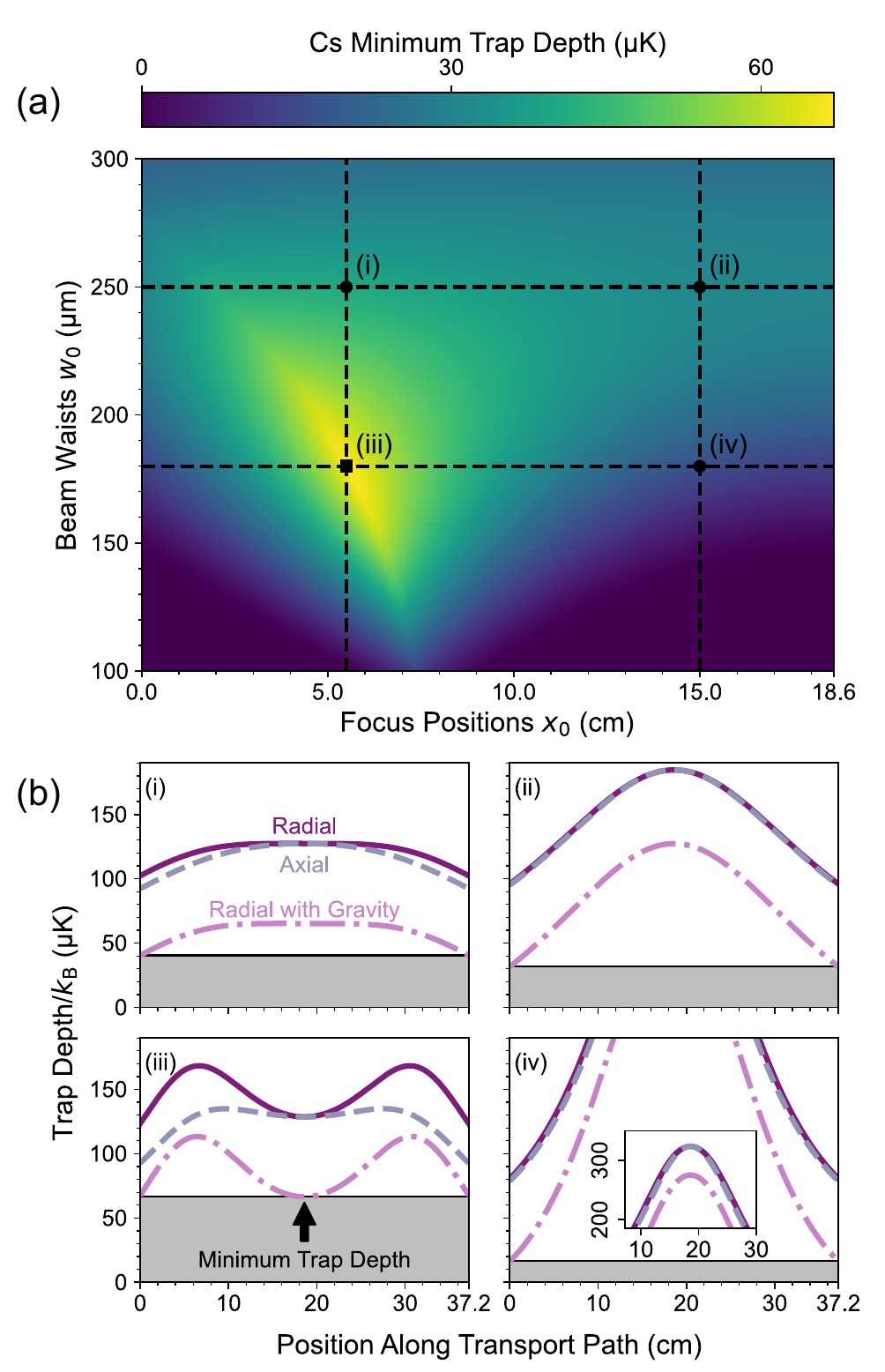"}
\caption{\label{fig:optimumBeams}
Optimization of the waists and focus positions of the lattice beams used for transport. (a) Contour plot of the minimum trap depth for Cs along the transport path as a function of the focus positions and waists of the beams. (b) Variation of the trap depth along the transport path in the axial (blue-gray dashed lines), radial (purple solid lines) and radial with gravity (magenta dash-dotted lines) directions for the different beam parameters indicated in (a). The minimum trap depth is indicated with the gray shaded region and is maximized in (iii) with $x_0 = 5.5$\,cm and $w_0= 180\,\mu$m, when the trap depth in the center is equal to that at either end.
}
\end{figure}

\begin{table}
\caption{\label{table:diffDistances} Comparison of the optimum beam parameters and trap depths achievable for different transport distances $d$ using $18$\,W per beam and $1064$\,nm. The vertical depth is limiting for all distances.}
\begin{ruledtabular}
\begin{tabular}{ccccc}
    $d$ (cm) & $x_0$ (cm) & $w_0$ ($\mu$m) & $\mathcal{U}_{\mathrm{Cs}}$ ($\mu$K) &  $\mathcal{U}_{\mathrm{Rb}}$ ($\mu$K) \\
    30   & 4.5 & 157 & 98 & 55 \\
    35   & 5.1 & 177 & 75 & 41 \\
    37.2 & 5.5 & 180 & 66 & 36 \\
    40   & 6.0 & 185 & 57 & 31 \\
\end{tabular}
\end{ruledtabular}
\end{table}

The theoretical framework outlined above allows the trap depth in each direction to be calculated as a function of distance along the transport axis and hence the optimal beam parameters for optical transport to be identified. In the experimental setup described in Sec.\,\ref{sec:expSetup} we use transport beams with equal powers. This symmetry means that we simply need to find $x_0$ and $w_0$, as defined in Fig.\,\ref{fig:radialComponents}(b). For the purposes of the calculations we use a conservative estimate of 18\,W in each beam. We note that whilst the choice of power scales the overall trap depth, it is not critical to the optimization. To perform the optimization, the trap depth along the entire transport path is calculated for each pair of $x_0$ and $w_0$ and the overall minimum is found. The optimal values for $x_0$ and $w_0$ are then those that maximize the minimum trap depth. 

Fig.\,\ref{fig:optimumBeams}(a) shows a contour plot of the minimum trap depth for Cs versus $x_0$ and $w_0$ with yellow/bright regions signifying deeper traps and blue/dark regions shallower traps. The focus position $x_0$ is varied from $0$\,cm to $18.6$\,cm. These limits correspond to the beams being focused at the ends or the center of the transport path, respectively. Fig.\,\ref{fig:optimumBeams}(b) shows how the trap depth varies along the transport path for the four different pairs of $x_0$ and $w_0$ values indicated in (a). The black horizontal line and the gray shaded region show the minimum trap depth in each case. As panel (iii) shows, the optimum beam parameters of $x_0 = 5.5$\,cm and $w_0 = 180\,\mu$m, balance the trap depths at the edges and at the center. 
Increasing the waist to $250\,\mu$m, as shown in panel (i), leads to a longer Rayleigh range which flattens the trap depth profile, but at the cost of reduced trap depth at the ends of the transport. In contrast, increasing $x_0$ to $15$\,cm, as shown in panel (iv), greatly increases the trap depth around the center of the transport path at the cost of a dramatic reduction at the ends. Finally, panel (ii) shows the case of larger $x_0$ and $w_0$ with a flatter profile than (iv) but still with the same limitation. In all cases, the trap depth is limited by effect of gravity. The same is true for Rb, although the lower polarizability of Rb at $1064$\,nm reduces the difference between the depths in the axial and vertical directions. Nevertheless, the optimum beam parameters are ultimately the same for the two species. The minimum trap depths achieved after this optimization are $66\,\mu$K for Cs and $36\,\mu$K for Rb.

The same optimization was repeated for several different transport distances and the results are summarized in Table\,\ref{table:diffDistances}. For longer transport distances, larger waists and longer Rayleigh ranges are required, as can be expected intuitively. The optimum focus positions also increase, moving further from the ends of the transport path. The minimum trap depth falls off with distance but still remains above $50\,\mu$K for Cs and above $30\,\mu$K for Rb at a transport distance of $40$\,cm.

\subsection{Magnetic levitation}

\begin{figure}
\includegraphics[width=\linewidth]{"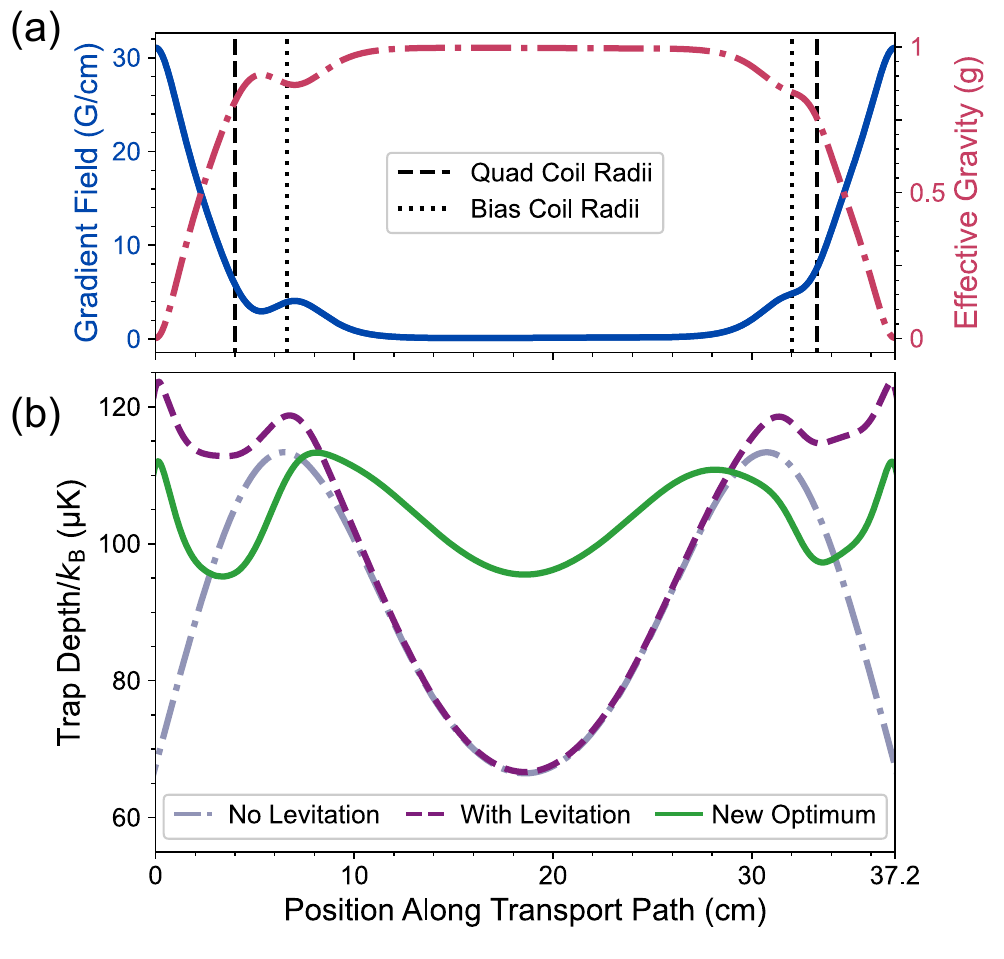"}
\caption{
\label{fig:levitation}
Simulated effect of adding magnetic levitation on the trap depth. 
(a) Effective range of the coils. Blue solid line shows the variation in the vertical gradient field along the transport path. Red dash-dotted lines shows the resultant effective gravitational force experiences by the Cs atoms. The radii of the quad (bias) coils are indicated by the vertical dashes (dotted) lines.
(b) Vertical radial trap depth for Cs along the transport path. Blue-gray dash-dotted line shows the previous, unlevitated case. The purple dashed line shows the trap depth with the addition of magnetic levitation and the same beam parameters. Re-optimizing taking the effect of levitation into account, yields beam parameters that give the trap depth shown by the green solid line.
}
\end{figure}

\begin{table}
\caption{\label{table:optima} Comparison of the optimum beam parameters and the minimum trap depths achieved for Rb and Cs with and without levitation using 18\,W per beam and 1064\,nm.}
\begin{ruledtabular}
\begin{tabular}{ccc}
     & Without Levitation & With Levitation\\
    $x_0$ (cm) & 5.5 & 7.2 \\
    $w_0$ ($\mu$m) & 180 & 195 \\
    \\
    Min radial $\mathcal{U}_{\mathrm{Cs}}$ ($\mu$K) & 66 & 95 \\
    Min radial $\mathcal{U}_{\mathrm{Rb}}$ ($\mu$K) & 36 & 54 \\
    \\
    Min axial $\mathcal{U}_{\mathrm{Cs}}$ ($\mu$K) & 93 & 93 \\
    Min axial $\mathcal{U}_{\mathrm{Rb}}$ ($\mu$K) & 55 & 55 \\
\end{tabular}
\end{ruledtabular}
\end{table}

Our experiment employs large volume dipole traps at the start and end of the transport that require magnetic levitation \cite{PhysRevA.63.023405}, and so it is natural to take the magnetic potential into account in the optimization of the beam parameters for transport. At low magnetic fields, the magnetic field gradient required to exactly levitate the atoms is given by
\begin{equation}
    \frac{d \vert \mathbf{B}\vert}{d z} = \frac{mg}{M_{\mathrm{F}} \vert g_{\mathrm{f}} \vert \mu_{\mathrm{B}}},
\end{equation}
where $g_{F}$ is the Land\'e $g$-factor. We work with Cs and Rb in their energetically lowest Zeeman states, $\ket{F = 3, m_{\mathrm{F}} = 3 }$ and $\ket{F = 1, m_{\mathrm{F}} = 1 }$, respectively. The magnetic field gradients required to fully levitate the two species in these states are almost identical, namely $30.6$\,G/cm for Rb and $31.1$\,G/cm for Cs, meaning no compromise between species needs to be found~\cite{lercherProductionDualspeciesBoseEinstein2011,McCarron2011}. 

In both chambers, the magnetic potentials are generated using a pair of coils in the anti-Helmholtz configuration to generate a quadrupole field, in combination with a pair of coils in the Helmholtz configuration to generate a bias field to vertically offset the field zero. The bias field is typically set to $22$G to minimize three-body loss of Cs~\cite{weberBoseEinsteinCondensationCesium2003}. This is important for other stages of the experiment, but is not critical to the transport. We set the magnetic field gradient to exactly levitate the atoms at the center of the MOT chamber and science cell. However, the magnitude of the gradient drops off rapidly along the transport path, as shown in Fig.\,\ref{fig:levitation}(a). Nevertheless, the levitation is still highly beneficial to the overall transport potential.

For both Cs and Rb, magnetic levitation approximately doubles the vertical trap depth at the edges of the transport path compared to the non-levitated case, as can been seen in Fig.\,\ref{fig:levitation}(b) by comparing the dash-dotted blue-gray line and the dashed purple line. Re-optimizing the beam parameters to maximize the minimum trap depth along the transport paths yields new values of $x_0 = 7.2$\,cm  and $w_0 = 195\,\mu$m, with the corresponding trap depth shown in Fig.\,\ref{fig:levitation}(b) by the solid green line. The change in the optimum parameters and the improvement in the transport potential are summarized in Table\,\ref{table:optima}. The minimum vertical trap depths are increased by $44$\,\% to $95\,\mu$K for Cs and by $50$\,\% to $54\,\mu$K for Rb, whilst the minimum axial depth remains unchanged at $93\,\mu$K for Cs and $55\,\mu$K for Rb.

\subsection{Trap Frequencies}
\label{sec:trapFrequency}
\begin{figure}
\includegraphics[width=\linewidth]{"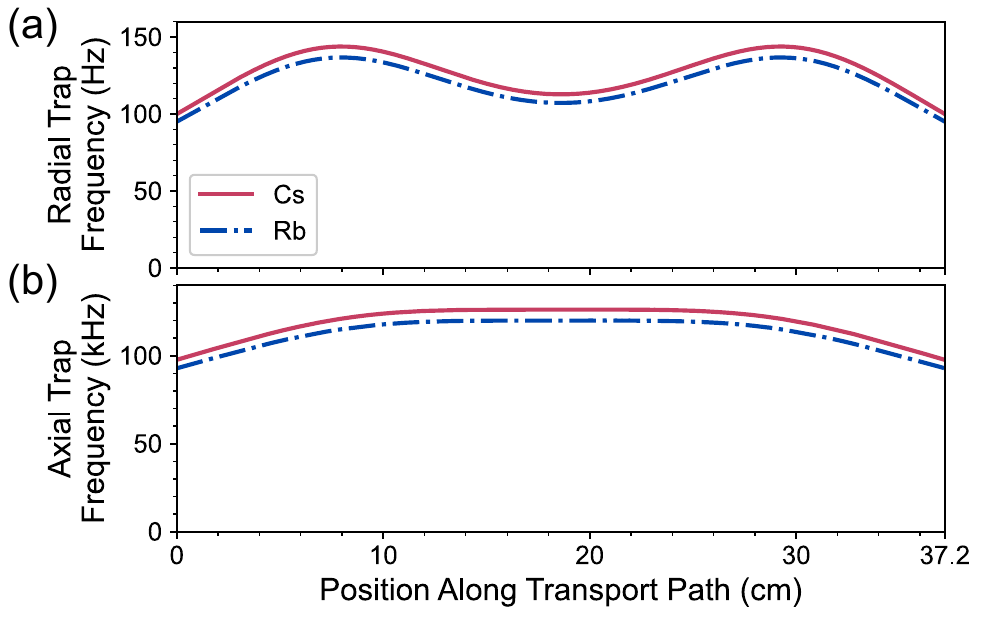"}
\caption{
\label{fig:trapFreq}
Variation of the predicted trap frequencies along the transport in the (a) radial direction and (b) axial direction. Cs is shown in the solid red line and Rb in the dash-dotted blue line. The calculations are presented for the optimum beam parameters with levitation,  $x_0 = 7.2$\,cm  and $w_0 = 195\,\mu$m, and 1064\,nm light with 18\,W per beam.}
\end{figure}

The predicted trap frequencies along the axis of the optical conveyor belt are shown in Fig.\,\ref{fig:trapFreq}, in both the axial and the radial directions. They are calculated for the optimum beam parameters with levitation, $x_0 = 7.2$\,cm and $w_0 = 195 \,\mu$m. The variation in trap frequency is approximately $\pm20$\,\% in the radial direction and $\pm10$\,\% axially. Note the high axial trap frequency of $\simeq110$\,kHz that allows fast acceleration of the atoms up to $38\,\mathrm{km/s^{2}}$. For higher accelerations, the axial trap depth is reduced to zero due to the accelerational tilting of the axial potential. We anticipate that changes in the trap frequency over the course of the transport could cause parametric heating~\cite{Greiner2001} if the time scale for the change approaches ${\left(2\nu_{\mathrm{rad}}\right)}^{-1} \approx 5$\,ms. This would lead to a drop of transport efficiency at very high speeds.

\section{Experimental Setup\label{sec:expSetup}}
\begin{figure*}
\includegraphics[width=\linewidth]{"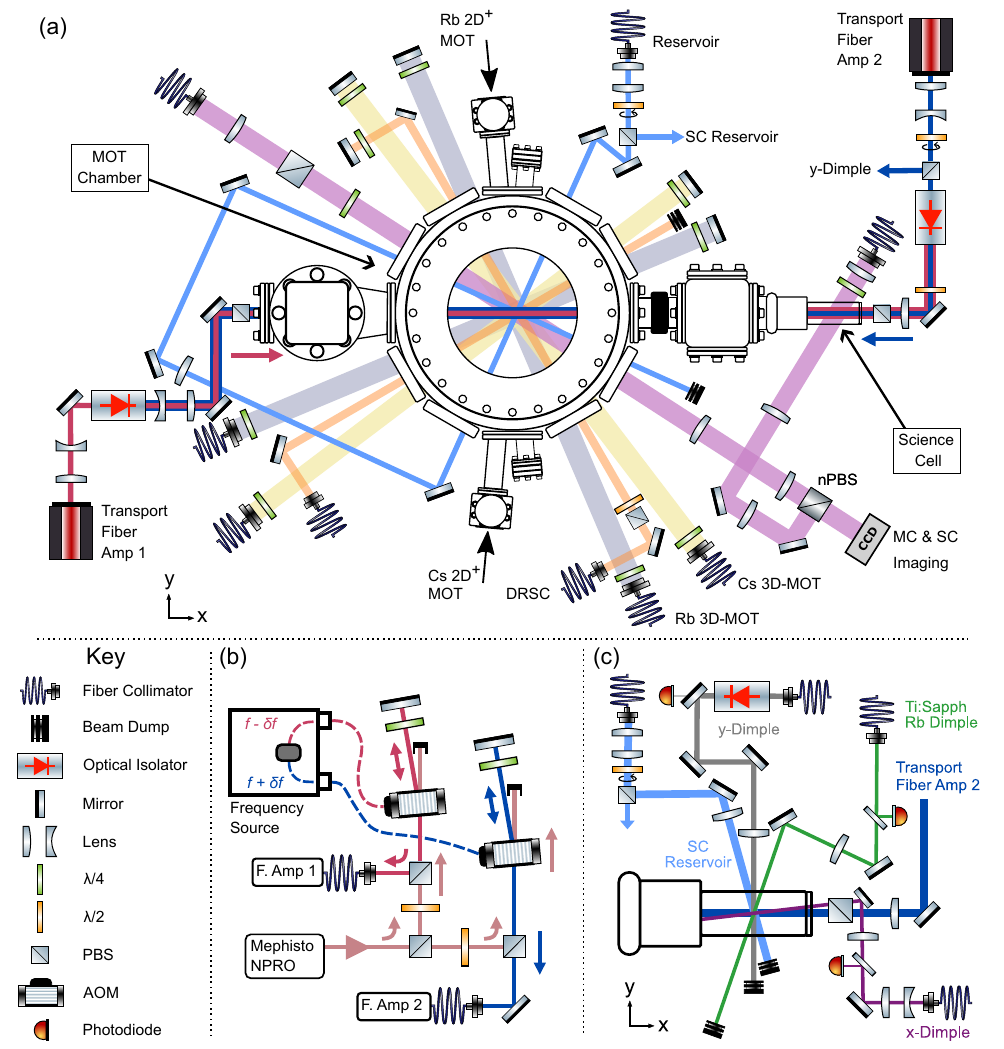"}
\caption{\label{fig:setupSketch}
Schematic overview of the experimental apparatus. 
(a) The vacuum apparatus and optical layout used to cool, trap, transport and image Rb and Cs atoms. The transport beams are aligned to be overlapped and counter-propagating along $x$, the axis between the two chambers. Electromagnetic coils used to generate bias and gradient magnetic fields sit above and below both chambers, but are omitted for clarity. Similarly, the MOT and DRSC beams propagating through the MOT chamber in the $z$ direction are also omitted.
(b) The optical setup of the seed path for the fiber amplifiers. Two double-passed AOMs are used to control the frequency difference between the light seeding the amplifiers. Different frequencies of light are shown in different colors. 
(c) The optical setup of the dipole traps in the science cell that are used to produce Bose-Einstein condensates of Cs or Rb. Photodiodes are used to servo the power in the dimple paths by controlling the RF power to an AOM before the respective optical fiber.
}
\end{figure*}

A schematic of the vacuum apparatus and the optical setup for cooling, trapping and transporting the atoms is shown in Fig.\,\ref{fig:setupSketch}. The initial cooling of Cs in our experiment, up to and including degenerate Raman sideband cooling (DRSC)~\cite{Kerman2000}, has been described in~\cite{Ratkata2021}. Rb is cooled using the same techniques. The experiment employs two separate $2\text{D}^+$ magneto-optical traps (MOTs)~\cite{Dieckmann1998}, one for Cs and one for Rb, leading to high-flux atomic beams that enter the MOT chamber on opposite sides. Atoms from these beams are captured in a 3D MOT in the center of the chamber. The atoms then undergo further sub-Doppler cooling using an optical molasses and degenerate Raman sideband cooling (DRSC)~\cite{Kerman2000} for both species simultaneously. This cools the atoms to $\sim1~\mu$K and spin-polarizes them into the energetically lowest Zeeman state, namely $\ket{F=3, m_{\text{F}}=3}$ for Cs and $\ket{F=1, m_{\text{F}}=1}$ for Rb. Subsequently the atoms are magnetically levitated and loaded into a large volume crossed optical dipole trap (the ``reservoir trap'') made from two beams with $\sim500\,\mu$m waists crossing at $\sim90$\,\degree. The beams are derived from a $50$\,W, $1070$\,nm broadband ytterbium fiber laser (IPG Photonics) and setup in a bow-tie configuration, wherein the light exiting the MOT chamber is reused for the second beam. Following a hold of 750\,ms to allow the atoms to thermalize in the reservoir trap, we typically have $1.0 \times 10^{7}$ Cs atoms at a temperature of $3.7\,\mu$K or $1.0 \times 10^{7}$ Rb atoms at a temperature of $5.6\,\mu$K.

The transport lattice is loaded from the reservoir trap.  We use a separate $30$\,W $1064$\,nm fiber amplifier (Azurlight Systems) for each of the two lattice beams. With the optical setup in Fig.\,\ref{fig:setupSketch}(a), we are able to deliver up to $20(1)$\,W to the atoms in each lattice beam. The power in each beam is controlled by a  $\lambda/2$ waveplate mounted onto a rotating, hollow-core stepper motor before a polarizing beam splitter (PBS)~\cite{Raghuram2023} and can be smoothly ramped on/off in $100$\,ms for transfer of atoms into/out of the lattice. The fiber amplifiers were chosen for their low relative intensity noise (RIN) at frequencies around the axial lattice frequencies. The use of long $\sim 1$\,m beam paths requires good pointing stability. After a warm-up period of up to two hours, the beam pointing out of the amplifiers was found to be stable within $4\,\mu$rad. Changes in the beam pointing were correlated with the cycle of the air conditioning in the lab.

The seed light for the amplifiers is derived from a $2$\,W, $1064$\,nm Mephisto-NPRO laser using the setup shown in Fig.\,\ref{fig:setupSketch}(b). The use of a double-passed acousto-optic modulator (AOM) in each seed path allows the frequency difference between the lattice beams to be controlled~\cite{Klostermann2022} 
up to a maximum of $\Delta f = 103$\,MHz~\footnote{A total frequency difference of $103$\,MHz is achievable when each AOM frequency is only shifted in one direction, positive (negative) for the AOM in the seed path for fiber amp. 1 (fiber amp. 2), as is done for one-way transport. When round-trip transport is performed, the alignment of the AOMs must be adjusted to achieve symmetric operation. Then a frequency difference of $\Delta f = \pm82$\,MHz is achievable}. 
Ensuring that the RF modulation applied to the AOMs does not introduce relative phase noise around the lattice trap frequencies is crucial to avoiding heating \cite{Klostermann2022,luNoiseSensitivitiesAtom2020}. Accordingly, we use a DDS-based dual-frequency generator (MOGLabs ARF) with phase synchronized outputs, and arbitrary frequency modulation at rates up to $1$\,MHz with a memory size of 8191 instructions. This memory size sets an upper bound on the length of ramps which can be implemented before the update rate becomes close to the axial trap frequency, which is known to cause heating~\cite{Hickman2020}.

The lenses for the transport beams were initially aligned to give the target beam profile, as measured by deflecting the beams onto a different path before the vacuum chamber. Typical uncertainties were $\sim 10\,\mu$m for the waist and $\sim 1$\,cm for the focus position. The beams out of the fiber amplifiers had slightly elliptical profiles and showed some astigmatism. As a consequence it was not possible to exactly match the desired beam sizes in both axes. Thus after the two transport beams were aligned and overlapped with each other, the lens positions were adjusted to maximize the measured transport efficiency. 

Following the transport to the science cell, the atoms are transferred into the optical dipole traps shown in Fig.\,\ref{fig:setupSketch}(c). Light from the 1070\,nm fiber laser is redirected to the science cell while the atoms are held in the transport lattice to form a new large volume ``SC reservoir'' trap. Similarly, light from transport fiber amplifier 2 is redirected to form the ``$y$-Dimple'' arm of a tighter crossed optical dipole trap. In both cases, the redirection of light is achieved in 50\,ms using the same hollow-core stepper motor setup used to switch the transport lattice light on/off~\cite{Raghuram2023}. The ``$x$-Dimple'' arm is derived directly from the Mephisto-NPRO used to seed the fiber amplifiers. An additional 830\,nm dimple trap derived from a Ti:sapphire laser is used for the trapping and cooling of Rb. Further details of the trapping in the science cell and evaporative cooling to form Bose-Einstein condensates are presented in Sec.\,\ref{sec:BEC}.

Throughout this work, we use absorption imaging to measure the number of atoms and their spatial distribution. This can be performed in either the MOT chamber or the science cell using the same CCD camera, as shown in Fig\,\ref{fig:setupSketch}(a).

\section{Transport Characterization\label{sec:transportCharacterize}}
We characterize and optimize the optical transport efficiency by comparing the number of atoms that arrive in the science cell with the number measured in the MOT chamber following a hold in a stationary lattice for the same duration as the transport. Evaluating the efficiency in this way compensates for the imperfect lattice loading, the lifetime of the atoms in the lattice and slow drifts in the initial atom number.

We first investigate different transport ramp profiles, comparing their efficiencies as a function of the maximum acceleration. We highlight the importance of the update rate of the RF frequency generator. We then explore various transport parameters, including the average transport speed and the power in the lattice beams.

\subsection{Transport ramp profiles}
\begin{figure}
\includegraphics[width=\linewidth]{"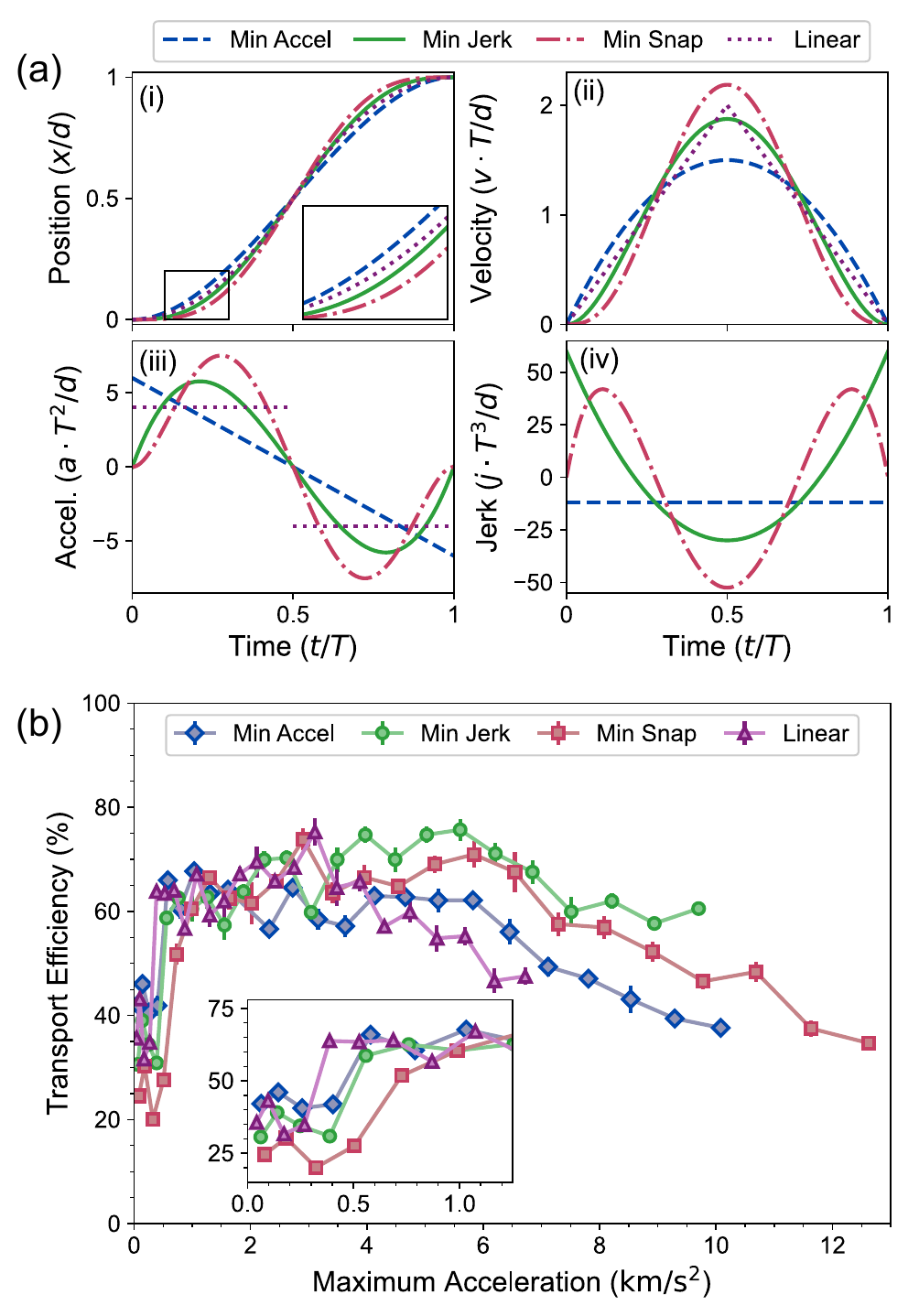"}
\caption{\label{fig:transportRamps}Comparison of transport ramps. (a) The variation with time of (i) position,  (ii) velocity, (iii)  acceleration and (iv) jerk for the four different ramp profiles: minimum acceleration (blue dashed lines), minimum jerk (green solid lines), minimum snap (red dash-dotted lines) and linear (purple dotted lines). All quantities are normalized to the distance of the ramp $d$ and the ramp duration $T$. The jerk for the linear profile is omitted but would consist of 3 Dirac-delta peaks at $t/T=$ $0$, $0.5$ and $1$. 
(b) The measured transport efficiency for Cs as a function of the maximum acceleration, for all four ramps. Error bars show the standard deviation of five repeat measurements. Lines connecting the points are included as a guide to the eye.
}
\end{figure}

We control the velocity profile of the atoms during transport by dynamically changing the frequency difference $\Delta f$ between the lattice beams, such that $v =  \lambda \Delta f /2$. We compare velocity profiles where the velocity varied linearly, to profiles designed to minimize the acceleration, jerk, and snap (rate of change of jerk) of the atoms during the transport.

The most commonly used trajectory is a ``linear'' ramp in which the velocity is increased linearly to a maximum value before being decreased at the same rate~\cite{Schrader2001, Langbecker2018, Klostermann2022, Bao2022}. This gives a trajectory with piece-wise constant acceleration such that the position is described by a pair of parabolas. Linear ramps can be preferable due to their comparative ease of implementation, often being an inbuilt feature of frequency control systems. However, the acceleration is discontinuously changed at the start, midpoint and end of a linear ramp. These sudden perturbations can lead to large losses of atoms, especially for large accelerations~\cite{Klostermann2022}. Alternative ramp profiles designed to minimize such perturbations have been used. These include sinusoidal ramps~\cite{Hickman2020}, ramps piece-wise cubic in acceleration~\cite{Schmid2006}, and the ``minimum jerk trajectory''~\cite{Liu2019}.  However with the exception of~\cite{Hickman2020} which investigated transport over a distance of $0.2$\,mm, little systematic investigation into the effectiveness of different ramp profiles has been reported. 

We use the minimum jerk trajectory which minimizes the square of the jerk integrated over the whole trajectory, following~\cite{Liu2019, Hogan1984}. The integral of the jerk is used as it is sensitive to perturbations throughout the entire ramp, and the square is used to avoid unwanted cancellations between points where the jerk has different signs. The function which minimizes this integral is a fifth-order polynomial:
\begin{equation}
\label{eqn:MJT}
	x (t) = d \left[10 {\left(\frac{t}{T}\right)}^3 - 15{\left(\frac{t}{T}\right)}^4 + 6 {\left(\frac{t}{T}\right)}^5 
	\right],
\end{equation}
where $d$ is the transport distance, $T$ is the transport duration and $t$ is time such that $ 0 \leq t  \leq T$.

We also investigate two additional trajectories, the ``minimum acceleration trajectory'' and the ``minimum snap trajectory''. These minimize the square integral of the acceleration and snap, respectively. Figure\,\ref{fig:transportRamps}(a) compares all the transport ramp profiles we investigate, presenting the position, velocity, acceleration and jerk as a function of the transport time. The exact functional form of each ramp is given in the Appendix. All four ramps can be completely described in terms of the transport duration $T$ and the transport distance $d$, which scale the $x$ and $y$ axes of the position plot, respectively. 

For each ramp we measured the transport efficiency to the science cell as a function of the ramp duration. To compare the performance of the ramps in Fig.\,\ref{fig:transportRamps}(b), the ramp duration is converted into a maximum acceleration, as described in the Appendix. The fastest ramp has a duration of $14.9$\,ms, limited by the diffraction efficiency bandwidth of the AOMs in the seed setup. Faster ramps could be used by increasing the power in the seed path. However, as can be seen in Fig.\,\ref{fig:transportRamps}(b), all ramps show a decline in transport efficiency for maximum accelerations above $6$\,km/$\rm{s}^2$. Optimum transport efficiencies of up to $75$\,\% are achieved between $2$\,km/$\rm{s}^2$ and $6$\,km/$\rm{s}^2$. The minimum jerk trajectory shows the clearest and broadest peak in efficiency between $3.5$\,km/$\rm{s}^2$ and $5.5$\,km/$\rm{s}^2$.

We attribute the decline in efficiency at high maximum accelerations to the reduction in the axial trap depth due to the accelerational tilt of the potential. For example, at $10$\,km/$\rm{s}^2$ the minimum axial trap depth is reduced by $32$\,\%, from $93\,\mu$K to $63\,\mu$K, for the minimum jerk trajectory~\footnote{The decrease in axial trap depth is ramp dependent as the maximum acceleration occurs at different positions for each ramp.}. For high accelerations, a clear difference between the ramps is evident with the minimum jerk trajectory performing the best, closely followed by the minimum snap trajectory. The minimum acceleration and linear trajectories both show substantial reductions in efficiency. A significant difference between the ramps is that the former two trajectories are continuous in acceleration, while the latter two have discontinuities in the acceleration (both have discontinuities at the start and end, but the linear ramp also has a discontinuity at $t = T/2$). Moreover, both the linear and minimum acceleration trajectories also require substantial acceleration at the ends of the transport where the axial trap depth is lowest, as shown in Fig.\,\ref{fig:optimumBeams}(b)(iii). These factors probably account for the differences in performance.

\begin{figure}
\includegraphics[width=0.96\linewidth]{"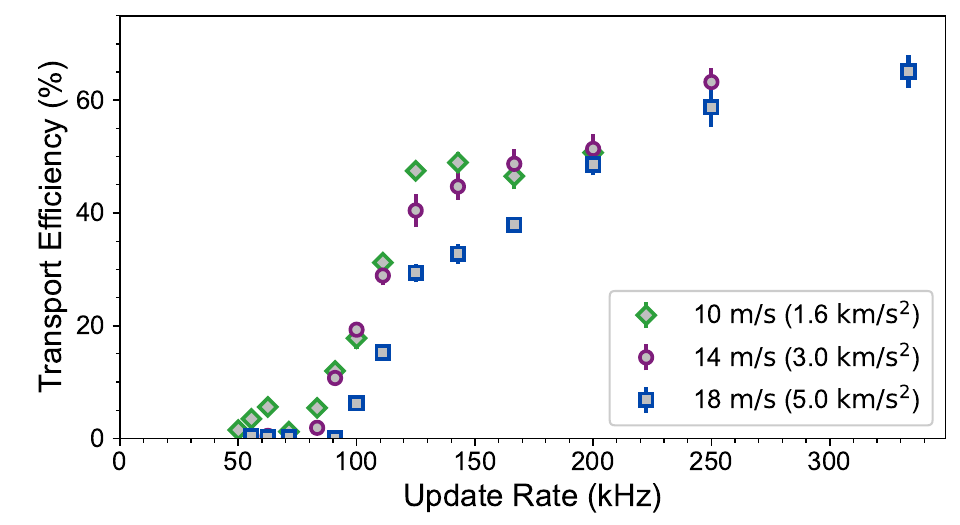"}
\caption{\label{fig:transport-update-rate} The dependence of the transport efficiency for Cs using the minimum jerk trajectory on the update rate of the RF generator for three different ramp speeds (maximum accelerations).}
\end{figure}

We attribute the sharp drop-off in efficiency for slower ramps to the limited update rate of the RF generator used to control the difference in the lattice frequencies. As the ramp speed decreases, the update frequency must also be decreased in order to be able to store the entire ramp in the limited memory of the RF generator. When the ramp duration is $\gtrsim 60$\,ms (equivalent to maximum accelerations $\lesssim 0.6$\,km/s$^2$ for the minimum jerk trajectory), the update frequency becomes comparable to the axial trap frequency, leading to parametric heating and loss~\cite{Hickman2020}. Note that the fastest usable update rate depends on the distance and duration of the ramp, but not on the trajectory used. In Fig.\,\ref{fig:transport-update-rate} we explicitly investigate the dependence of the efficiency of transporting Cs to the science cell on the update rate. Here we use only the minimum jerk trajectory. The figure highlights the importance of using update rates significantly above the axial trap frequency ($\simeq110\,$kHz in this case). As an example, we see that for a maximum acceleration of $5\,\mathrm{km/s^2}$, reducing the update rate from $333$\,kHz to $167$\,kHz leads to a drop in efficiency from $66$\,\%  to $38$\,\%.

To crudely probe the energy distribution of the transported atoms, we also measured the number of atoms loaded into a crossed optical dipole trap with a depth that was $\sim 50$\,\% lower than the transport lattice. We found that the dependence on the maximum acceleration remained broadly the same as that shown in Fig.\,\ref{fig:transportRamps}(b). Although, above $6$\,km/s$^2$ the difference in performance of the minimum jerk and minimum snap trajectories became indiscernible. This suggests that the additional atoms transported by the minimum jerk trajectory evident in Fig.\,\ref{fig:transportRamps}(b) are in the high energy tail of the distribution. 
Similarly, the performance of the minimum acceleration and linear trajectories also became the same above $6$\,km/s$^2$, but still at a substantially reduced level compared to the other two trajectories. This further highlights the importance of a continuous acceleration profile.

Based upon this study, we conclude the minimum jerk trajectory is optimal in our application. In the remainder of this section, we therefore exclusively use this trajectory to further characterize the transport.

\subsection{Transport speed}
\begin{figure}
\includegraphics[width=\linewidth]{"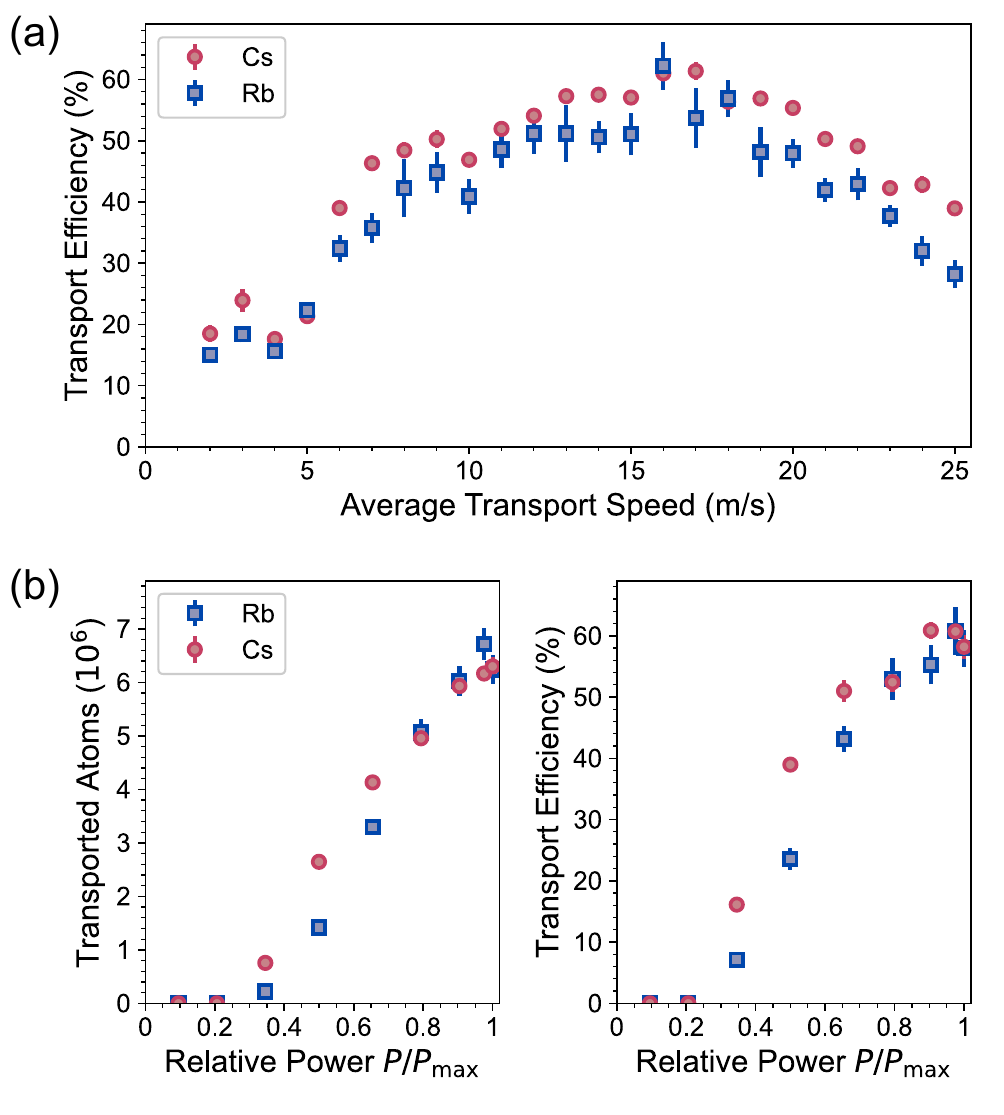"}
\caption{\label{fig:oneSpeciesCharacter}
Characterization of single species transport for Cs (red circles) and Rb (blue squares) separately using the minimum jerk trajectory. (a) The dependence of the transport efficiency on the average transport speed. (b) The effect of varying the power in the transport lattice beams on the number of atoms that reach the science cell (left) and the transport efficiency (right). For these data $P_{\rm{max}}=20(1)$\,W in each beam.
}
\end{figure}

Figure\,\ref{fig:oneSpeciesCharacter}(a) shows the dependence of the transport efficiency on the average transport speed, now for both Cs and Rb. The Rb data shows much the same behavior as Cs, with the drop-off in efficiency at low speeds due to the update rate and the decline at high speeds due to accelerational tilting. Despite the significantly lower polarizability, and hence trap depth, for Rb the achievable transport efficiency is only marginally worse than for Cs. Indeed, both species have a maximum in transport efficiency of 50-60\,\%~\footnote{The slight decrease in efficiency compared with Fig.\,\ref{fig:transportRamps} resulted from realignment of the transport lattice following an issue with one of the fiber amplifiers that led to a change in the initial loading of the lattice.} for speeds around $16$\,m/s, equivalent to a transport duration of $23.35$\,ms. We therefore use this speed for the subsequent measurements.

\subsection{Transport lattice power}
Figure\,\ref{fig:oneSpeciesCharacter}(b) shows both the number of atoms transported and the transport efficiency for Rb and Cs as a function of the power used to form the lattice, up to a maximum of $P_{\rm{max}}=20(1)$\,W in each beam. Both the atom number and the efficiency increase approximately linearly above a threshold of $P/P_{\rm{max}}\simeq0.3$, although the efficiency begins to roll off for $P/P_{\rm{max}}\gtrsim0.8$. It follows that for $P/P_{\rm{max}}\gtrsim0.8$ the number of atoms transported is limited by the number loaded into the transport lattice rather than insufficient trap depth leading to loss during the transport. The data also indicate that more atoms could be delivered to the science cell by further increasing the laser power, for example, by using higher power fiber amplifiers.

\section{Evaporation to Bose-Einstein Condensation}\label{sec:BEC}
\begin{figure}
\includegraphics[width=\linewidth]{"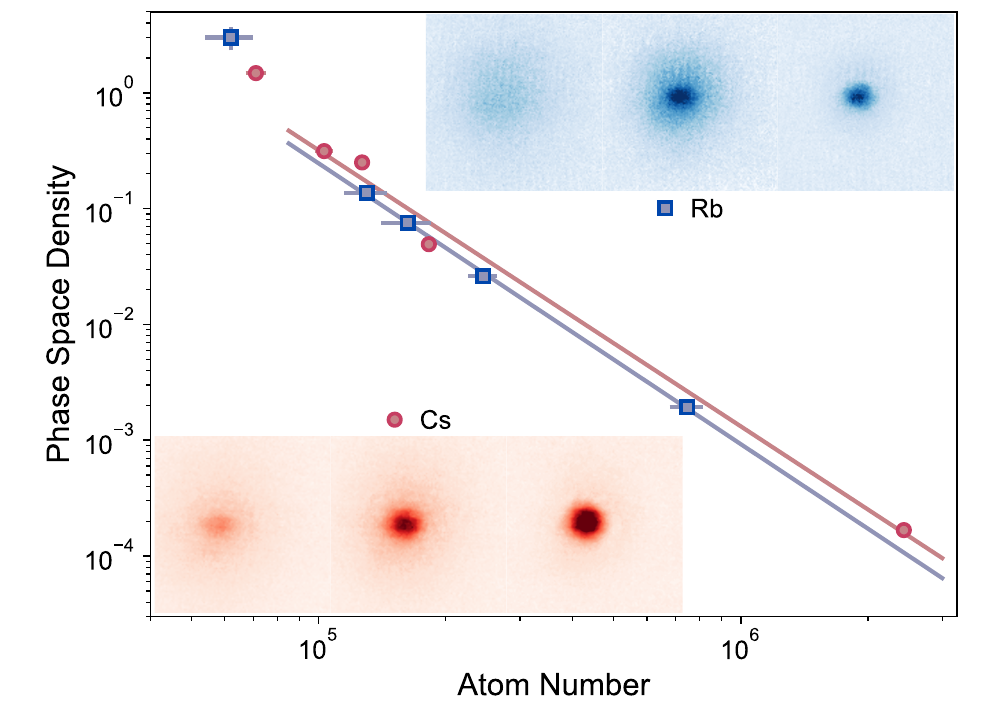"}
\caption{\label{fig:becEvaporationtrajectory}
Evaporation trajectories to Bose-Einstein condensation (BEC) for Cs (red circles) and Rb (blue squares). Phase space density is plotted as a function of the atom number. Linear fits yield evaporation efficiencies of $2.4(5)$ for Rb and $2.38(15)$ for Cs. The insets show, from left to right, optical depth images of the clouds as they are cooled through the BEC transition. 
}
\end{figure}

Following the transport to the science cell, the atoms are transferred into the optical dipole traps shown in Fig.\,\ref{fig:setupSketch}(c) and further cooled evaporatively to reach the Bose-Einstein condensation transition. We use the established approach for Cs, based upon initial loading into a large volume reservoir trap, followed by transfer into a tighter dimple trap~\cite{weberBoseEinsteinCondensationCesium2003}. A similar approach is employed for Rb. 

The reservoir trap is formed from two beams. The lattice beam focused closest to the science cell is retained, with the other beam being ramped off to remove the lattice. Confinement along the transport axis is provided by an elliptical $1064$\,nm beam with a horizontal waist of $510\,\mu\text{m}$, a vertical waist of $200\,\mu$m and a maximum power of $40$\,W. Magnetic levitation, using a gradient of 31\,G/cm and a bias field of 40\,G, compensates for gravity.  The combined trap is roughly mode matched to the elongated shape of the cloud following transport. Around  $2.3(1) \times 10 ^6$ Cs and  $2.1(1) \times 10 ^6$ Rb are captured in this trap, at temperatures ${\simeq5}\,\mu$K. 

The dimple trap is formed by a pair of more tightly focused $1064$\,nm beams; the $x$-dimple has a waist of $45\,\mu$m and the $y$-dimple has a waist of $100\,\mu$m. These relatively large dimple waists are chosen to limit the three-body loss for Cs in the final stages of evaporation \cite{weberBoseEinsteinCondensationCesium2003}. To avoid forming accidental lattices, all $1064$\,nm beams used in evaporation are mutually detuned by at least $80$\,MHz.  

The dimple trap is loaded by ramping up both beams to depth to around $10\,\mu$K in $100$\,ms. Effective loading of the dimple trap requires a high collision rate in the gas and fast thermalization. For Cs this is achieved via the high the s-wave length of $\simeq870\,a_0$ at the applied bias field of 40\,G. In the Rb sequence an additional $50\,\mu$m dipole trap at $830$\,nm is used to achieve higher trap frequencies and, hence, an increased collision rate.

Forced evaporation proceeds by first ramping off the reservoir trap in $2$\,s. For Rb the $830$\,nm beam is also ramped off in this step to reduce heating from near-resonant photon scattering. Then the power in the dimple is ramped down in two steps, each lasting 2\,s, while under-levitating the cloud to tilt the trap in the vertical direction. During the final step the tilt of the trap is increased to maintain sufficiently high trap frequencies and collision rate for efficient cooling~\cite{hungAcceleratingEvaporativeCooling2008}. For each step we optimize the levitation field and dipole trap powers to maximize the evaporation efficiency. Crucially for Cs it is important to lower the magnetic field to the 3-body loss minimum~\cite{weberBoseEinsteinCondensationCesium2003} at 22\,G during the final two steps of evaporation.

Figure\,\ref{fig:becEvaporationtrajectory} shows the evaporation trajectory to BEC for both Rb and Cs independently. The atom number and temperature are measured using absorption imaging following time-of-flight expansion. The trap frequencies are calibrated by measuring center-of-mass oscillations for various trap powers. These quantities are combined to calculate the phase space density of the gases. We extract evaporation efficiencies~\footnote{The evaporation efficiency is $\gamma=(d\ln\rho)/d\ln N)$, where $\rho$ is the phase space density and $N$ the number of atoms.} of $2.4(5)$ for Rb and $2.38(15)$ for Cs from fits to the data in the non-degenerate regime. 
The BEC transition is typically crossed with around $4\times 10^4$ Cs atoms or $5\times 10^4$ Rb atoms. The insets to Fig.\,\ref{fig:becEvaporationtrajectory} show representative absorption images taken after 45\,ms time-of-flight expansion. The ability to produce pure Bose-Einstein condensates in the science cell validates the choice of transport scheme.

\section{Dual-species transport}\label{sec:DualSpecies}

\begin{figure}
\includegraphics[width=\linewidth]{"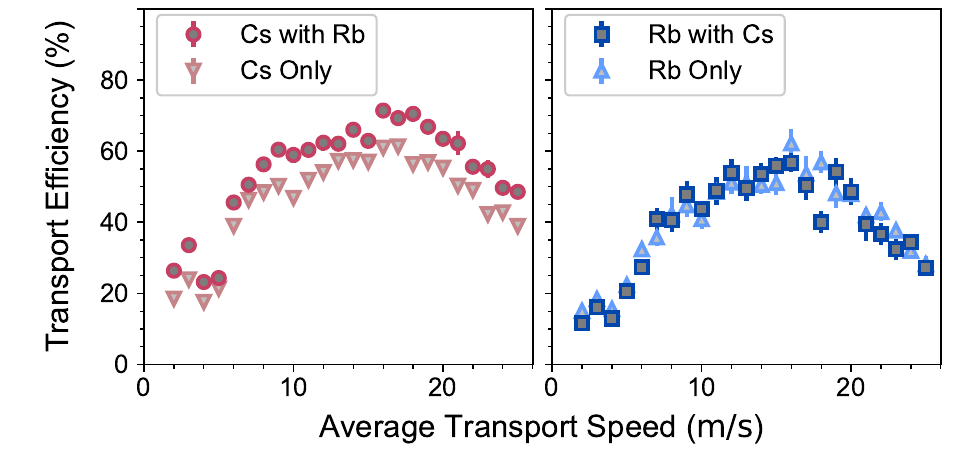"}
\caption{\label{fig:dualSpeciesCharacter} Dual-species transport of Rb and Cs. The dependence of the efficiency on the average speed is presented for both dual-species and single-species transport. Different laser-cooling routines are used for the dual-species and single species measurements resulting in different initial atom numbers (see text for details).
}
\end{figure}

Finally, in Fig.\,\ref{fig:dualSpeciesCharacter} we demonstrate the simultaneous dual-species transport of Cs and Rb. For comparison results are also presented without the second species. All the measurements were taken on the same day for consistency. We see that the transport efficiency of both Rb and Cs is not negatively affected by the presence of the other species. Indeed for Cs there is actually a slight increase in the transport efficiency. This probably results from thermalization with Rb prior to and during loading as a consequence of the large interspecies scattering length of $\sim650\,a_0$~\cite{Takekoshi2012}; the Rb gas experiences shallower traps than Cs throughout and is therefore expected to be colder. However, the interpretation is complicated by the fact that the routine we use to simultaneously cool and load both species results in significantly fewer atoms loaded into the transport lattice initially. 

For the dual-species routine, the initial cooling stages in the MOT chamber (the MOT, the compressed MOT, molasses and DRSC) are implemented simultaneously and therefore must share some parameters. Crucially, the timing of each stage can no longer be independently optimized for the two species and a compromise must be found. In particular, for the DRSC stage we find that a duration of 15\,ms is optimum for Rb alone, whereas 5\,ms is optimum for Cs alone. For the measurements shown in Fig.\,\ref{fig:dualSpeciesCharacter} we used a routine that favored Rb, leading to a $\sim50$\,\% reduction in the number of Cs atoms loaded into the lattice from $\sim1.0\times10^7$ to $\sim5\times10^6$, whereas the number of Rb atoms was $9\times10^6$ compared to $\sim1.0\times10^7$ for the single species routine. The dramatic reduction in the number of Cs atoms may also contribute to the apparent slight increase in efficiency if the smaller cloud is better matched to the transport lattice resulting in a colder distribution and, hence, less loss during the transport.

Modifications to the laser systems used for DRSC will allow better simultaneous cooling of Rb and Cs in the MOT chamber, therefore enabling the transport of sufficient atoms to the science cell to produce dual-species BECs following a separated trapping scheme~\cite{Reichsoellner2017}. Nevertheless, our current results clearly demonstrate that efficient dual-species transport of Rb and Cs is possible using a relatively simple optical conveyor-belt setup.

\section{Conclusion\label{sec:conclusion}}
We have demonstrated fast, efficient optical transport of both cesium and rubidium atoms using an optical conveyor belt. By carefully choosing the waists and focus positions of the two lattice beams, we are able to use simple Gaussian beams rather than employing more complicated Bessel beams or a variable-focus lens. Transport was performed and characterized for both Cs and Rb, with up to $7\times 10^6$ atoms of either species being transported over $37.2$\,cm in under $25$\,ms. We compared different transport trajectories, finding the performance of the minimum jerk trajectory to be the best. To demonstrate the viability of our transport scheme
for further experiments in the science cell, the production of BECs of either Rb or Cs was demonstrated. Finally dual-species transport was demonstrated, without any additional losses compared to single-species transport. 

Fast and efficient dual-species transport of Rb and Cs lays a groundwork for the study of RbCs molecules in a quantum gas microscope. Implementing separate optical traps for Rb and Cs in the science cell will facilitate the production of dual-species condensates~\cite{lercherProductionDualspeciesBoseEinstein2011}. The protocol for the preparation of heteronuclear atom pairs in a 3D optical lattice is established~\cite{Reichsoellner2017} and a compatible association sequence for the production of ground-state molecules has recently been demonstrated~\cite{Das2023}. Single site resolved imaging of the molecules may be achieved by reversing the association sequence and detecting the resulting atoms using standard atomic quantum gas microscopy techniques~\cite{Rosenberg2022,Christakis2023}. RbCs molecules are especially attractive in this regard as microscopy has been demonstrated for both atomic species using simple cooling techniques~\cite{Bakr2009,Impertro2023}. Moreover, magic wavelength trapping has been demonstrated for rotational states in RbCs molecules~\cite{Guan2021,Gregory2023} leading to second-scale rotational coherence and new opportunities for the simulation of models relating to quantum magnetism~\cite{Barnett2006,CapogrossoSansone2010,Hazzard2013,Blackmore2018}.

\section*{Rights retention statement}
For the purpose of open access, the authors have applied a Creative Commons Attribution (CC BY) license
to any Author Accepted Manuscript version arising from
this submission.

\section*{Data availability statement}
The data presented in this work are available from
Durham University~\cite{data}.

\section*{Acknowledgments}
The authors would like to thank Till Klostermann and Annie Jihyun Park for fruitful discussions. This work was supported by UK Engineering and Physical Sciences Research Council (EPSRC) Grant EP/P01058X/1, UK Research and Innovation (UKRI) Frontier Research Grant EP/X023354/1, the Royal Society and Durham University.

\bibliography{transport}

\clearpage

\appendix*

\section{Transport ramp equations}\label{appendix:Ramps}
The equation for the four different transport ramps are given below. As for the main text, the equations are written in terms of the transport distance $d$ and transport duration $T$. 

\noindent Minimum acceleration trajectory:
\begin{equation}
	x (t) =  d \left[3 {\left(\frac{t}{T}\right)}^2 - 2{\left(\frac{t}{T}\right)}^3
	\right]
\end{equation}
Minimum jerk trajectory:
\begin{equation}
	x (t) = d \left[10 {\left(\frac{t}{T}\right)}^3 - 15{\left(\frac{t}{T}\right)}^4 + 6 {\left(\frac{t}{T}\right)}^5 
	\right]
\end{equation}
Minimum snap trajectory:
\begin{equation}
\begin{aligned}
    x (t) = 
    & d \Biggl[35 {\left(\frac{t}{T}\right)}^4 - 84{\left(\frac{t}{T}\right)}^5 \\
    & + 70 {\left(\frac{t}{T}\right)}^6 - 20{\left(\frac{t}{T}\right)}^7
	\Biggr]
\end{aligned}
\end{equation}
Linear trajectory:
\begin{equation}
	x (t) = 
	\begin{cases}
		d \left[ 2 {\left(\frac{t}{T}\right)}^2 \right] & \text{if } 0 \leq t/T \leq \frac{1}{2} \\
		&\\
		d \left[ -1 + 4 {\left(\frac{t}{T}\right)} - 2{\left(\frac{t}{T}\right)}^2 \right]& \text{if } \frac{1}{2} < t/T \leq 1
	\end{cases}
\end{equation}

For each ramp profile the maximum acceleration can be calculated using
\begin{equation}
 a_{\text{max}} = \alpha \frac{d}{T^{2}}
\end{equation}
where $a_{\text{max}}$ is the maximum acceleration and $\alpha$ is a constant depending on the trajectory used, with $\alpha = 6$ for minimum acceleration; $\alpha = \nicefrac{10}{\sqrt{3}}$ for minimum jerk; $\alpha = \nicefrac{84}{5\sqrt{5}}$ for minimum snap and $\alpha = 4$ for linear.

\end{document}